\documentclass[journal]{IEEEtran}
\usepackage{graphicx}
\usepackage{amsmath}
\usepackage{amssymb}
\usepackage{epic}
\usepackage{fullpage}

\usepackage{graphics}
\usepackage{epsfig}
\usepackage{subfigure}
\usepackage{latexsym}
\usepackage{cite}

\allowdisplaybreaks

\newcommand{\qed}{\hspace*{\fill} $\Box$ \\}
\newcommand{\bs}{\boldsymbol}

\def\ba{\begin{array}}
\def\ea{\end{array}}
\def\be{\begin{equation}}
\def\ee{\end{equation}}
\def\bea{\begin{eqnarray}}
\def\eea{\end{eqnarray}}
\def\beas{\begin{eqnarray*}}
\def\eeas{\end{eqnarray*}}

\newtheorem{theorem}{Theorem}

\newtheorem{definition}{Definition}
\newtheorem{lemma}{Lemma}

\begin{document}

\title{Throughput Optimal Distributed Control of Stochastic Wireless Networks}
\author{Yufang Xi and Edmund M. Yeh\\
    Department of Electrical Engineering\\
    Yale University\\
    New Haven, CT 06520, USA\\
    {\tt \{yufang.xi,edmund.yeh\}@yale.edu}}

\maketitle

\setcounter{footnote}{1} \footnotetext{This research is supported
in part by Army Research Office (ARO) Young Investigator Program
(YIP) grant DAAD19-03-1-0229 and by National Science Foundation
(NSF) grant CCR-0313183.}

\begin{abstract}
The Maximum Differential Backlog (MDB) control policy of Tassiulas and Ephremides has been shown to
adaptively maximize the stable throughput of multi-hop wireless networks with random traffic arrivals and
queueing.  The practical implementation of the MDB policy in wireless networks with mutually interfering
links, however, requires the development of distributed optimization algorithms.  Within the context of
CDMA-based multi-hop wireless networks, we develop a set of node-based scaled gradient projection power
control algorithms which solves the MDB optimization problem in a distributed manner using low communication
overhead.  As these algorithms require time to converge to a neighborhood of the optimum, the optimal rates
determined by the MDB policy can only be found iteratively over time.  For this, we show that the iterative
MDB policy with convergence time remains throughput optimal.
\end{abstract}


\section{Introduction}\label{sec:Introduction}

The optimal control of multi-hop wireless  networks is a major
research and design challenge due, in part, to the interference
between nodes, the time-varying nature of the communication
channels, the energy limitation of mobile nodes, and the lack of
centralized coordination.  This problem is further complicated by
the fact that data traffic in wireless networks often arrive at
random instants into network buffers.  Although a complete
solution to the optimal control problem is still elusive, a major
advance is made in the seminal work of Tassiulas and
Ephremides~\cite{paper:TE92}. In this work, the authors consider a
stochastic multi-hop wireless network with random traffic arrivals
and queueing, where the activation of links satisfies specified
constraints reflecting, for instance, channel interference.  For
this network, the authors characterize the stability region, i.e.
the set of all end-to-end demands that the network can support.
Moreover, they obtain a {\em throughput optimal} routing and link
activation policy which stabilizes the network whenever the
arrival rates are in the interior of the stability region, without
{\em a priori} knowledge of arrival statistics.  The throughput
optimal policy operates on the Maximum Differential Backlog (MDB)
principle, which essentially seeks to achieve load-balancing in
the network.  The MDB policy (sometimes called the ``backpressure
algorithm") has been extended to multi-hop networks with general
capacity constraints in~\cite{paper:NMR03} and has been combined
with congestion control mechanisms
in~\cite{paper:NMR05,paper:ES05}.

While the MDB policy represents a  remarkable achievement, there remains a significant difficulty in applying
the policy to wireless networks.  The mutual interference between wireless links imply that the evaluation of
the MDB policy involves a centralized network optimization.  This, however, is highly undesirable in wireless
networks with limited transmission range and scarce battery resources.  The call for distributed scheduling
algorithms with guaranteed throughput gives rise to two main lines of research.

One approach is to adopt simple physical and MAC layer models and apply computationally efficient scheduling
rules in a distributed manner. The work in \cite{paper:CKS05,paper:WSP06} study networks where at any instant
only mutually non-interfering links are activated, and any link, as long as it is active, transmits at a
fixed rate. In particular, it is shown in~\cite{paper:CKS05} that Maximal Greedy Scheduling can achieve a
guaranteed fraction of the maximum throughput region.  This result is generalized in~\cite{paper:WSP06} to
multi-hop networks where the end-to-end paths are given and fixed. Despite its simplicity, the distributed
scheduling considered in the above work applies to only a limited class of networks. Moreover, the simplicity
is gained at the expense of throughput optimality~\cite{paper:LS05}.

Another line of research develops distributed power control and rate allocation algorithms for implementing
the MDB policy with the aim of preserving the throughput optimality. Thus far, distributed MDB control has
been investigated only for networks with relatively simple physical layer models. For example,
Neely~\cite{paper:Nee05} studies a cell partitioned network model where different cells do not interference
with each other so that scheduling can be decentralized to each cell. However, the question of how the MDB
policy can be efficiently applied in general wireless networks remains unanswered.

In this paper, we consider the  implementation of the MDB algorithm within interference-limited CDMA wireless
networks, where transmission on any given link potentially contends with interference from all other active
links.  In this setting, we present two main sets of results.  First, we develop a set of node-based scaled
gradient projection power control algorithms which solves the MDB optimization in a distributed manner using
low communication overhead.  As these algorithms require time to converge to a neighborhood of the optimum,
the optimal rate allocation of the MDB policy can only be found iteratively over time.  In the second result,
we show that the iterative MDB policy with convergence time remains throughput optimal as long as the second
moments of the traffic arrival rates are bounded. Combining these two results, we conclude that our
algorithms yield a distributed solution to throughput optimal control of CDMA wireless networks with random
traffic arrivals. The framework and techniques developed in this work can readily be adapted to general
interference-limited stochastic wireless networks.

Iterative implementation of the MDB policy has also been studied
independently by Giannoulis et al.~\cite{paper:GTT06}. They
investigate distributed power control algorithms for CDMA networks
which are iterated once for every update of the queue state. In
their scheme, the power and rate allocation algorithms are
iterated only once in a slot, after which the queue state is
re-sampled. In contrast, our algorithms re-sample the queue state
only after the power and rate allocation has converged to the
optimum for the previous queue state.  We provide a rigorous proof
for the throughput optimality of our scheme using a novel
geometric method. We also compare the performance of our scheme
with that of the non-convergent iterative algorithms
in~\cite{paper:GTT06} and of the centralized MDB policy (whereby
it is assumed that the optimal powers and rates are
instantaneously obtained given the queue state) through
simulations. In the experiments we conducted, the iterative MDB
policy with convergence outperforms that without convergence.

This paper is organized as follows. Section~\ref{sec:Formulation} introduces the model of stochastic
multi-hop wireless networks and reviews the throughput optimal Maximum Differential Backlog policy. The
node-based power control algorithms that achieve the optimal rate allocation are presented in
Section~\ref{sec:DMDB}. In Section~\ref{sec:StabilityOfAsyn}, we prove that the iterative MDB policy proposed
in the last section maintains the throughput optimality even in the presence of non-negligible convergence
time. The comparison of different implementations of the MDB policy is conducted through simulations in
Section~\ref{sec:Simulation}.

\section{Network Model and \\Throughput Optimal Control}\label{sec:Formulation}

\subsection{Model of Stochastic Multi-hop Wireless Networks}

Consider a wireless network represented by a directed and connected graph $\mathcal G = ( {\cal N}, {\cal E}
)$. Each node $i \in {\cal N}$ models a wireless transceiver. An edge $(i,j) \in {\cal E}$ represents a
unidirectional  radio channel from node $i$ to $j$. For convenience, let ${\cal O}(i) \triangleq \{j:(i,j)
\in {\cal E}\}$ and ${\cal I}(i) \triangleq \{j:(j,i) \in {\cal E}\}$ denote the sets of node $i$'s next-hop
and previous-hop neighbors, respectively. Let the vector $\bs h = (h_{ij})_{(i,j) \in {\cal E}}$ represent
the (constant) channel gains on all links.

Denote the transmission power used on link $(i,j)$ at (continuous) time $\tau$ by $P_{ij}(\tau) \ge 0$, and the
instantaneous service rate of link $(i,j)$ by $R_{ij}(\tau) \ge 0$. A feasible service rate vector $\bs R(\tau) =
(R_{ij}(\tau))_{(i,j) \in {\cal E}}$ must belong to a given {\em instantaneous feasible rate region} ${\cal C}(\bs
P(\tau))$ reflecting the physical-layer coding mechanism. Under peak power constraints $\hat P_i, i \in {\cal N}$, let
\[\Pi = \left\{\bs P(\tau) \in \mathbb{R}_+^{|{\cal E}|}:~\sum_{j \in {\cal O}(i)} P_{ij}(\tau) \le \hat P_i, \forall i
\in {\cal N} \right\}\] be the set of feasible power allocations and \[{\cal C}(\Pi) \triangleq {\rm conv}\left(
\bigcup_{\bs P \in \Pi} {\cal C}(\bs P)\right)\] be the {\em long-term feasible service rate region}. Here, the convex
hull operation ${\rm conv}(\cdot)$ indicates the possibility of time
sharing among different feasible power allocations $\bs P \in \Pi$ over a sufficiently long period. 

Let the data traffic in the network  be classified according to their destinations. Traffic of type $k \in
{\cal K}$ is destined for a set of nodes ${\cal N}_k \subset {\cal N}$ (when type $k$ traffic reaches any
node in ${\cal N}_k$, it exits the network), where ${\cal K}$ is the set of all traffic types. Let $T > 0$ be
a given time slot length. Let the number of bits of type $k$ entering the network at node $i$ from time $tT$
to $(t+1)T$ be a nonnegative random variable $B_i^k[t]$.  Assume that for all $t \in \mathbb{Z}_+$,
$B_i^k[t]$ are independent and identically distributed with $\mathbb{P}(B_i^k[t] = 0) > 0$. Let $\mathbb{E}
B_i^k[t] = a_i^k < \infty$ and $\mathbb{E} \left(B_i^k[t]\right)^2 = b_i^k < \infty$ be the first and second
moments of $B_i^k[t]$. Furthermore, assume all arrival processes $\{B_i^k[t]\}_{t=1}^\infty$, $i \in {\cal
N}, k \in {\cal K}$ are mutually independent.

Assume node $i \in {\cal N}$ provides a (separate) infinite buffer
$i^k$ for each type $k$ of traffic that is not destined for $i$.
Denote the unfinished work in $i^k$ at time $\tau$ by
$U_i^k(\tau)$. We focus on the queue states sampled at slot
boundaries $\tau = tT$, $t \in \mathbb{Z}_+$. Let $U_i^k[t]$
denote the instantaneous backlog at the beginning of the $t$th
slot, i.e., $U_i^k[t] = U_i^k(tT)$.  Over the $t$th slot, link
$(i,j)$ serves $i^k$ at rate $R_{ij}^k[t] = \int_{tT}^{(t+1)T}
R_{ij}^k(\tau) d \tau$. The aggregate service rate on link $(i,j)$
over the $t$th slot is $R_{ij}[t] = \sum_{k \in {\cal K}}
R_{ij}^k[t]$. Thus, we have the following queueing dynamics:
\be\label{eq:QueueDynamic}
\begin{array}{l}
U_i^k[t+1] \le \\
\left( U_i^k[t] - \displaystyle\sum_{j \in {\cal O}(i)}R_{ij}^k[t]
+ \displaystyle\sum_{m \in {\cal I}(i)}R_{mi}^k[t]  +
B_i^k[t]\right)^+. \end{array}  \ee Here $(x)^+$ denotes
$\max\{x,0\}$, and the inequality comes from the fact that in
general, since certain queues may be empty, the actual endogenous
arrival rate is less than or equal to the nominal rate $\sum_{m
\in {\cal I}(i)}R_{mi}^k[t]$.

\subsection{Stability Region and Throughput Optimal Policy}

Given the wireless network model, we now define notions of
stability  and investigate throughput optimal control policies.

\vspace{0.1in} \begin{definition}\cite{paper:NMR03}\label{def:QueueStable} The queue $i^k$ is \emph{stable} if
$g_i^k(\xi) \triangleq \lim\sup_{n\rightarrow \infty} \frac{1}{n}\sum_{t = 1}^{n} \mathbb{P}\left[U_i^k[t]
> \xi \right]  \rightarrow 0$ as $\xi \rightarrow \infty$.
Input processes $\{\bs B[t] = (\bs B_i^k[t])_{i \in {\cal N}, k \in
{\cal K}}\}_{t=1}^\infty$ are \emph{stabilizable} if there exist
service processes $\{R_{ij}^k[t]\}$ for all $(i,j)\in {\cal E}$
and $k \in {\cal K}$ such that for every $t \in \mathbb{Z}_+$,
$\bs R[t] \in {\cal C}(\Pi)$,\footnote{Here we assume the slot
length $T$ is long enough for time-sharing among different $\bs P
\in \Pi$.} and the resulting queueing processes are all stable.
\end{definition}

\vspace{0.1in}
\begin{definition}
The \emph{stability region} $\Lambda$ of a wireless multi-hop network is the closure of the set of the
average arrival rate vectors $\bs a$ of all stabilizable input processes.
\end{definition}

\vspace{0.1in}

For a general wireless multi-hop network, its stability region has a simple characterization in terms of
supporting multi-commodity rates that are feasible under link capacity constraints.

\vspace{0.1in}\begin{theorem}\cite{paper:NMR03}\label{thm:NetworkStabRegion} The stability region $\Lambda$ of the
wireless multi-hop network with transmission power constraint $\Pi$ is the set of all average rate vectors $(a_i^k)$
such that there exists a multi-commodity flow vector $(f_{ij}^k)$ satisfying
\[
f_{ij}^k \ge 0, \quad \forall (i,j)\in {\cal E}~\textrm{and}~k \in {\cal K},
\]
\[
a_i^k \le \sum_{j \in {\cal O}(i)} f_{ij}^k - \sum_{m \in {\cal I}(i)} f_{mi}^k, \quad \forall i \in {\cal N},~ k \in
{\cal K},
\]
\[
\sum_{k \in {\cal K}} f_{ij}^k \le C_{ij}, \; \forall (i,j)\in {\cal E}~\textrm{where}~(C_{ij})_{(i,j)\in{\cal E}} \in
{\cal C}(\Pi).
\]
\end{theorem}\vspace{0.1in}

The following Maximum Differential Backlog (MDB) policy has been
shown to be {\em throughput optimal} \cite{paper:TE92,paper:NMR03}
in the sense that it stabilizes all input processes with average
rate vectors belonging to the interior of $\Lambda$, without
knowledge of arrival statistics. The policy can be described as
follows:

\begin{enumerate}
\item At slot $t$, find traffic type $k^*_{ij}[t]$ having the
\emph{maximum differential backlog} over link $(i,j)$ for all
$(i,j) \in {\cal E}$. That is, $ k^*_{ij}[t] = \arg\max_{k \in
{\cal K}} \left\{U_i^k[t] - U_j^k[t]\right\}$, where $U_j^k[t]
\equiv 0$ if $j \in {\cal N}_k$. Let $b_{ij}^*[t] = \max\left\{0,
U_i^{k^*}[t] - U_j^{k^*}[t]\right\}$, where $k^* \equiv
k^*_{ij}[t]$.

\item Find the rate vector $\bs R^*[t]$ which solves
\be\label{eq:OptPowerVector} \max_{\bs R \in {\cal C}(\Pi)}
\sum_{(i,j) \in {\cal E}} b_{ij}^*[t] \cdot R_{ij}. \ee

\item The service rate provided by link $(i,j)$ to queue $i^k$ is
determined by
\[
R_{ij}^k[t] = \left\{\begin{array}{ll}
                        R_{ij}^*[t], \quad &\text{if}~k =
                        k^*_{ij}[t], \\
                        0, &\text{otherwise}.
                        \end{array} \right.
\]
\end{enumerate}

 For wired networks, the above MDB policy can be implemented in
a fully distributed manner. In wireless networks, however, the capacity of a link is usually affected by
interference from other links. Therefore, solving \eqref{eq:OptPowerVector} in general requires centralized
computation.  Thus far, distributed solutions for \eqref{eq:OptPowerVector} are available only for relatively
simple physical layer models~\cite{paper:Nee05}.

In the following, we develop efficient distributed MDB control algorithms for interference-limited CDMA
networks with random traffic. Throughout the rest of the paper, we assume all nodes have synchronized clocks
so that the boundaries of time slots at all nodes are aligned. This assumption guarantees that the MDB values
in~\eqref{eq:OptPowerVector} are taken at the same instant across all links. The study of MDB policy based on
asynchronously sampled queue state will be a subject of future work.

\section{Distributed Maximum Differential Backlog Control}\label{sec:DMDB}

\subsection{Throughput Optimal Power Control}

We study a wireless network using direct-sequence spread-spectrum CDMA. The received
signal-to-interference-plus-noise ratio (SINR) per channel code symbol of link $(i,j)$ is given by
\[
SINR_{ij} = \frac{Kh_{ij} P_{ij}}{\theta_i h_{ij}(P_i-P_{ij}) + \sum\limits_{m \ne i}{h_{mj}P_m} + N_j},
\]
where $K$ is the processing gain, $P_m = \sum_{k \in {\cal
O}(m)}{P_{mk} }$ is the total transmission power of node $m$, and
$N_j$ represents the noise power of receiver $j$. The parameter
$\theta_i \in [0,1]$ characterizes the degree of
self-interference.\footnote{$\theta_i = 0$ corresponds to the case
when node $i$ applies mutually orthogonal direct sequences for
transmissions to its receivers. In this case, signals intended for
different receivers will not interfere with each other in
demodulation. The other extreme, where $\theta_i = 1$, represents
the most pessimistic case where self-interference is as
significant as all other sources of interference.}

Assume the receiver of every link decodes its own signal against the interference from other links as
Gaussian noise. The information-theoretic capacity of link $(i,j)$ is given by
\[
R_s \log \left(1 + \displaystyle\frac{Kh_{ij}P_{ij}}{\theta_i h_{ij}(P_i - P_{ij}) + \sum\limits_{m \ne
i}{h_{mj}P_m} + N_j }\right).
\]
For convenience, we normalize the channel symbol rate $R_s$ to be one for subsequent analysis. We
also take $\log(\cdot)$ to be the natural logarithm to simplify differentiation operations.

In most CDMA systems, due to the large multiplication factor $K$, the SINR \emph{per symbol}
\[\frac{K h_{ij}P_{ij}}{\theta_i h_{ij}(P_i - P_{ij}) + \sum\limits_{m \ne i}{h_{mj}P_m} + N_j }\] is
typically high \cite{book:TV04}. Therefore, in the high SINR regime, we can approximate the capacity of any
active link $(i,j)$ by \[ \log\left(\displaystyle\frac{Kh_{ij}P_{ij}}{\theta_i h_{ij} \sum_{k \ne j}{P_{ik}}
+ \sum\limits_{m \ne i}{h_{mj} \sum_{k \in {\cal O}(m)}{P_{mk}}} + N_j }\right).
\]

With a change of variables $S_i = \ln P_i$, $\hat S_i = \ln \hat P_i$, and $S_{ik} = \ln P_{ik}$, the
capacity function becomes \beas && C_{ij}(\boldsymbol S) = \log(Kh_{ij}) + S_{ij} - \nonumber \\
&& \log\left(\theta_i h_{ij} \sum_{k \ne j}{e^{S_{ik}}} + \sum_{m \ne i}{h_{mj} \sum_{k \in {\cal
O}(m)}{e^{S_{mk}}}} + N_j \right), \label{eq:CapacityTransformed} \eeas which is known to be concave in $\bs
S$ \cite{paper:JXB03,paper:Chi04}. It follows that the instantaneous achievable region $\bigcup_{\bs P \in
\Pi} {\cal C}(\bs P)$ is convex, and therefore is equal to ${\cal C}(\Pi) = {\rm conv}\left( \bigcup_{\bs P
\in \Pi} {\cal C}(\bs P)\right)$.

Thus, the optimization problem in \eqref{eq:OptPowerVector} at a fixed time slot can be seen as optimizing over the
region\footnote{Notice that even if $\bigcup_{\bs P \in \Pi} {\cal C}(\bs P)$ is not convex, restricting the feasible
set of the optimization in~\eqref{eq:OptPowerVector} to $\bigcup_{\bs P \in \Pi} {\cal C}(\bs P)$ does not lose any
optimality. This is because the objective function is linear in the link rates, and so the maximum over any compact
region region is equal to the maximum over the convex hull of that region.} $\bigcup_{\bs P \in \Pi} {\cal C}(\bs P)$.
More specifically, it can be rewritten as the following concave maximization problem \bea
\textrm{maximize} && \sum_{(i,j)\in{\cal E}} b_{ij}^* R_{ij} \label{eq:BasicMDB} \\
\textrm{subject to} && R_{ij} = C_{ij}(\bs S), \;
\forall (i,j) \in {\cal E}, \nonumber \\
&& \sum_{j \in {\cal O}(i)} e^{S_{ij}} \le \hat P_i, \quad \forall i \in {\cal N}. \nonumber \eea Without
loss of generality, we assume $b_{ij}^* > 0$ for all $(i,j)$ (otherwise we can simply exclude those links
having $b_{ij}^* = 0$ from the objective function in \eqref{eq:BasicMDB}).

\subsection{Power Adjustment Variables}

\begin{figure}[h]
\begin{center}
\includegraphics[height = 4cm]{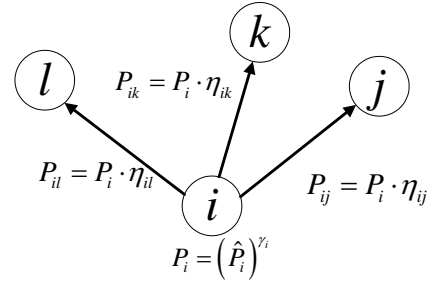}
\caption{Transmission powers in terms of the power control and power allocation
variables.}\label{fig:PowerCtrl}
\end{center}
\end{figure}

Next we introduce a set of node-based control variables for adjusting the transmission powers on all links.
They are \beas \textrm{Power allocation variables:} && \eta _{ik} \triangleq \frac{P_{ik}}{P_i}, \;
(i,k) \in {\cal E}, \label{eq:PowerAllocVariable} \\
\textrm{Power control variables:} && \gamma_i \triangleq \frac{S_i}{\hat S_i}, \; i \in {\cal N}.
\label{eq:PowerCtrlVariable} \eeas These variables are illustrated in Figure \ref{fig:PowerCtrl}.
With appropriate scaling, we can always let $\hat P_i
> 1$ for all $i \in {\cal N}$ so that $\hat S_i
> 0$. Therefore, we have the following equivalent Throughput Optimal Power Control (TOPC) problem:

\vspace{0.1in}maximize \be \label{eq:SubProblem} \sum_{(i,j)} b_{ij}^* \log  \frac{K h_{ij}(\hat P_i)^{\gamma_i}
\eta_{ij}}{\theta_i h_{ij}(\hat P_i)^{\gamma_i}(1-\eta_{ij}) + \displaystyle\sum_{m \ne i} h_{mj}(\hat P_m)^{\gamma_m}
+ N_j} \ee
 \beas \textrm{subject to} && \eta_{ij} \ge 0, \quad \forall (i,j) \in {\cal E}, \nonumber \\
                    && \sum_{j \in {\cal O}(i)} \eta_{ij} = 1, ~ \gamma_i \le 1, \quad \forall i \in {\cal N}. \eeas

\subsection{\label{sec:OptimalityConditions}Conditions for Optimality}

To solve the TOPC problem in \eqref{eq:SubProblem}, we compute the gradients of the objective function,
denoted by $F$, with respect to the power allocation variables and the power control variables, respectively.
They are as follows. For all $i \in {\cal N}$ and $j \in {\cal O}(i)$, \beas \frac{\partial F}{\partial
\eta_{ij}} &=& P_i \left[ \sum_{k \in {\cal O}(i)} b_{ik}^*
\frac{-\theta_i h_{ik}}{IN_{ik}} \right. \\
&& \left. - \sum_{m \ne i}\sum_{k \in {\cal O}(m)} b_{mk}^* \frac{h_{ik}}{IN_{mk}} + \delta\eta_{ij}\right],
\eeas where the \emph{power allocation marginal gain indicator} is
\begin{equation}\label{eq:MarginalPowerAlloc}
\delta\eta_{ij} \triangleq b_{ij}^*\left(\frac{1}{P_{ij}}+\frac{\theta_i h_{ij}}{IN_{ij}}\right).
\end{equation}
For all $i \in {\cal N}$,
\[ \frac{\partial F}{\partial \gamma_{i}} = \hat S_i \cdot \delta\gamma_i,\]
where the \emph{power control marginal gain indicator} is \bea \delta\gamma_i &\triangleq& P_i \left[\sum_{m
\ne i}\sum_{k \in {\cal O}(m)}  \frac{-b_{mk}^* h_{ik}}{IN_{mk}} + \right. \nonumber \\
&& \left.  \sum_{k \in {\cal O}(i)}
 \frac{-\theta_i b_{ik}^* h_{ik}}{IN_{ik}} + \sum_{k \in {\cal O}(i)} \delta\eta_{ik}\cdot \eta_{ik}\right]. \label{eq:MarginalPowerCtrl}
 \eea
The term $IN_{ij}$ appearing above is short-hand notation for the overall interference-plus-noise power at
the receiver end of link $(i,j)$, that is
\[
IN_{ij} = \theta_i h_{ij} \displaystyle\sum_{k \ne j}{e^{S_{ik}}} + \sum\limits_{m \ne i}{h_{mj} \sum_{k \in
{\cal O}(m)}{e^{S_{mk}}}} + N_j.
\]

The marginal gain indicators fully characterize the optimality conditions as follows.

\vspace{0.1in}\begin{theorem}\label{thm:OptimalityCondition} A feasible set of transmission power variables
$\{\eta_{ik}\}_{(i,k) \in {\cal E}}$ and $\{\gamma_i\}_{i \in {\cal N}}$ is the solution of the TOPC problem
\eqref{eq:SubProblem} if and only if the following conditions hold. For all $i \in {\cal N}$, there exists a
constant $\nu_i$ such that
\begin{eqnarray}
\delta\eta_{ik} = \nu_i, &\;& \forall k \in {\cal O}(i), \label{eq:PowerOptCond1}
\\
\delta\gamma_i = 0, &\;& \textrm{if}~\gamma_i < 1, \label{eq:PowerOptCond2}
\\
\vspace{1mm} \delta\gamma_i \ge 0, &\;& \textrm{if}~\gamma_i = 1. \label{eq:PowerOptCond3}
\end{eqnarray}
Here, all $\eta_{ik} > 0$ since $b_{ik}^* > 0$ by assumption.
\end{theorem}\vspace{0.1in}

For the detailed proof of Theorem~\ref{thm:OptimalityCondition}, see~\cite{paper:XY06}.  Due to the distributed form of
the optimality conditions, every node can check the conditions with respect to its controlled variables locally, and
adjust them towards the optimum. In the next section, we present a set of distributed algorithms that achieve the
globally optimal power configuration.

\subsection{\label{sec:Algorithms}Distributed Power Control
Algorithms}

We design scaled gradient projection algorithms which iteratively
update the nodes' power allocation variables and power control
variables in a distributed manner, so as to asymptotically
converge to the optimal solution of~(\ref{eq:SubProblem}). At each
iteration, the variables are updated in the positive gradient
direction, scaled by a positive definite matrix. When an update
leads to a point outside the feasible set, the point is projected
back into the feasible set~\cite{book:Ber99}.

\vspace{0.1in}
\subsubsection{Power Allocation Algorithm (PA)}

At the $k$th iteration at node $i$, the current local power allocation vector $\boldsymbol\eta_i^k =
(\eta_{ij}^k)_{j \in {\cal O}(i)}$ is updated by
\[
\bs\eta_i^{k+1} = PA(\bs\eta_i^k) = \left[ \bs\eta_i^k + \beta_i^k \cdot (Q_i^k)^{-1} \cdot \delta\bs\eta_i^k
\right]_{Q_i^k}^+.
\]
Here, $\delta\boldsymbol\eta_i^k = (\delta\eta_{ij}^k)_{j \in {\cal O}(i)}$ and $\beta_i^k$ is a positive
stepsize.  The matrix $Q_i^k$ is symmetric, positive definite on the subspace $\{ \boldsymbol v_i : \sum_{j
\in {\cal O}(i)}{v_{ij}} = 0 \}$. Finally, $[\cdot]_{Q_i^k}^+$ denotes the projection on the feasible set of
$\bs\eta_i$ relative to the norm induced by $Q_i^k$.\footnote{In general, $[\tilde{\bs x}]_{Q_i^k}^+ \equiv
\arg\min_{\bs x \in {\cal F}} (\bs x - \tilde{\bs x})' \cdot Q_i^k \cdot (\bs x - \tilde{\bs x})$, where
${\cal F}$ is the feasible set of $\bs x$.}

Suppose each node $j$ can measure the value of $SINR_{ij}$ for any of its incoming links. Before an iteration of $PA$,
node $i$ collects the current $SINR_{ij}$'s via feedback from its next-hop neighbors $j$. Node $i$ can then readily
compute all $\delta\eta_{ij}$'s according to
\[
\delta\eta_{ij} = b_{ij}^*\left(\frac{1}{P_{ij}}+\frac{\theta_i h_{ij}}{IN_{ij}}\right) =
\frac{b_{ij}^*}{P_{ij}}\left(1+\frac{\theta_i SINR_{ij}}{K}\right).
\]
Note that since the calculation of $\delta\eta_{ij}$ \emph{involves only locally obtainable measures}, the
$PA$ algorithm does not require global exchange of control messages.

\vspace{0.1in}
\subsubsection{Power Control Algorithm (PC)}

After a phase for exchanging control messages (which will be discussed below), every node $i$ is able to
calculate its power control marginal gain indicator $\delta\gamma_i$. From a network-wide viewpoint, the
power control vector $\bs\gamma^k = (\gamma_i^k)_{i \in {\cal N}}$ is updated by
\[
\bs\gamma^{k+1} = PC(\bs\gamma^k) = \left[ \bs\gamma^k + \xi^k \cdot (V^k)^{-1} \cdot \delta\bs\gamma^k
\right]_{V^k}^+.
\]
Here, $\xi^k$ is a positive stepsize and matrix $V^k$ is symmetric and positive definite. Note that
$PC$ becomes amenable to distributed implementation if and only if $V^k$ is diagonal.

We now derive an efficient protocol which allows each node to calculate its own $\delta\gamma_i$ given
limited control messaging. We first re-order the summations on the RHS of \eqref{eq:MarginalPowerCtrl} as
\begin{small}\beas &&\delta\gamma_i = P_i \left[  \sum_{j \ne i}\left\{-h_{ij}\sum_{m\in {\cal
I}(j)}{\frac{b_{mj}^*}{IN_{mj}}}\right\} \right. \\
&& \left. \sum_{j \in {\cal O}(i)}\left\{b_{ij}^*\left[\frac{1}{P_i} + (\theta_i\eta_{ij} - \theta_i +
1)\frac{h_{ij}}{IN_{ij}}\right] \right\} \right]. \eeas\end{small} With reference to the above expression, we
propose the following procedure for computing $\delta\gamma_i$.

\begin{figure}[h]
\begin{center}
\includegraphics[height = 5cm]{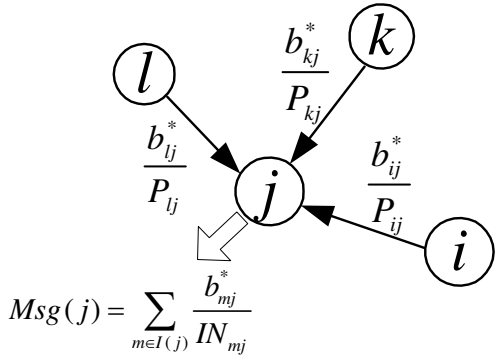}
\caption{Information Exchange Protocol for Power Control Algorithm}\label{fig:PC}
\end{center}
\end{figure}

\vspace{0.1in}\emph{Power Control Message Exchange Protocol:} Let each node $j$ assemble the measures
$\frac{b_{mj}^*}{IN_{mj}}$ from all its incoming links $(m,j)$. For this purpose, an upstream neighbor $m$ needs to
inform $j$ of the value $b_{mj}^* / P_{mj}$. Since node $j$ can measure both $SINR_{mj}$ and $h_{mj}$, it can calculate
\[ \frac{b_{mj}^*}{IN_{mj}} = \frac{b_{mj}^*}{P_{mj}} \frac{SINR_{mj}}{h_{mj} K}.
\] After obtaining the measures from all incoming links, node $j$ sums them up to form the power control message:
\[Msg(j) \triangleq \sum_{m\in {\cal I}(j)}{\frac{b_{mj}^* }{IN_{mj}}}.\] It then broadcasts $Msg(j)$ to the
whole network. The process for control messaging is illustrated by Figure \ref{fig:PC}, where the solid arrows
represent local message communication and the hollow arrow signifies the broadcasting of the message.

Upon obtaining $Msg(j)$ from node $j \ne i$, node $i$ processes it according to the following rule. If $j$ is a
next-hop neighbor of $i$, it multiplies the message with $h_{ij}$ and subtracts the product from the local measure
\beas &&b_{ij}^* \left[\frac{1}{P_i} + (\theta_i\eta_{ij} - \theta_i + 1)\frac{h_{ij}}{IN_{ij}}\right]\\ &=&
\delta\eta_{ij} \cdot \eta_{ij} + \left( \delta\eta_{ij} - \frac{b_{ij}^*}{P_{ij}}\right)
\frac{1-\theta_i}{\theta_i}.\eeas Otherwise, it multiplies $Msg(j)$ with $-h_{ij}$. Finally, node $i$ adds up the
results derived from processing all other nodes' messages, and this sum multiplied by $P_i$ equals $\delta\gamma_i$.
Note that in a symmetric duplex channel, 
 $h_{ij} \approx h_{ji}$, and node $i$ may use its own measure of $h_{ji}$ in place of $h_{ij}$. Otherwise, it will need
channel feedback from node $j$ to calculate $h_{ij}$. To summarize, the protocol requires {\em only one message from
each node} to be broadcast to the whole network.

\vspace{0.1in}
\subsubsection{Convergence of Algorithms}

\label{sec:ConvergenceProof}

We now formally state the central convergence result for the $PA$ and $PC$ algorithms discussed above.

\vspace{0.1in}\begin{theorem}\label{thm:AlgorithmConvergence} From any feasible initial transmission power
configuration $\{\boldsymbol\eta_i^0\}$ and $\boldsymbol\gamma^0$, there exist valid scaling matrices $\{Q_i^k\}$ and
$V^k$, and positive stepsizes $\{\beta_i^k\}$ and $\xi^k$ such that the sequences generated by the algorithms $PA(
\cdot )$ and $PC( \cdot )$ converge, i.e., $\boldsymbol\eta_i^k \to \boldsymbol\eta_i^*$ for all $i$, and
$\boldsymbol\gamma^k \to \boldsymbol\gamma^*$ as $k \to \infty$. Furthermore, $\{\boldsymbol\eta_i^*\}$ and
$\boldsymbol\gamma^*$ constitute a set of jointly optimal solution to the TOPC problem \eqref{eq:SubProblem}.
\end{theorem}\vspace{0.1in}

In the $PA$ and $PC$ algorithms, the scaling matrices are chosen
to be appropriate diagonal matrices which approximate the relevant
Hessians such that the objective value is increased by every
iteration until the optimum is achieved. This allows the scaled
gradient projection algorithms to approximate constrained Newton
algorithms, which are known to have fast convergence rates.
Furthermore, the scaling matrices are shown to be easily
calculated at each node using very limited control messaging.  The
detailed derivation of these parameters and the full proof of
Theorem~\ref{thm:AlgorithmConvergence} can be found in
\cite{paper:XY06}.

Also note that the convergence of the algorithms does not require any particular order of running $PA$ and
$PC$ algorithms at different nodes. Any node $i$ only needs to update its own variables $\bs\eta_i$ and
$\gamma_i$ using $PA$ and $PC$ until its local variables satisfy the optimality conditions
\eqref{eq:PowerOptCond1}-\eqref{eq:PowerOptCond3}.

\section{Throughput Optimality of Iterative Maximum Differential Backlog Policy}\label{sec:StabilityOfAsyn}

Since the $PA$ and $PC$ algorithms need a certain number of
iterations before reaching a close neighborhood of an optimum to
the problem in~\eqref{eq:SubProblem}, the optimal service rates
dictated by the MDB policy cannot be applied instantaneously.
Rather the optimal service rates can only be found iteratively
over time. At any moment in the convergence interval, the queues
are served at the rates which are iteratively updated towards the
optimal rates for the queue state at the beginning of a slot. The
service rates obtained at the end of a convergence period are
optimal only for the queue state some time ago. The effect of
using lagging optimal service rates is studied in the context of
$N \times N$ packet switches by Neely et al.
\cite{paper:NMR02}\footnote{In~\cite{paper:NMR02}, the current
queue state is taken to be the state of the Markov chain used for
stability analysis.  As we show below, however, the Markov state
should consist of the current queue state as well as the previous
queue state.} and in a queueing network with Poisson arrivals and
exponential service rates by Tassiulas and
Ephremides~\cite{paper:TE96}.  In \cite{paper:NMR02,paper:TE96},
however, the process of finding the optimal rates is not
iterative. It is assumed that once the (outdated) queue state
information becomes available, the optimal rates are obtained
instantaneously. Here, we analyze the iterative MDB algorithm with
convergence time in general multi-hop networks with i.i.d. random
arrival processes and general rate regions. We show that the
throughput optimality of the MDB policy is preserved for any
finite convergence time. For this, we invent a new geometric
approach for computing the expected Lyapunov drift of the queue
state.

\subsection{Transient Optimal Rates}

Without loss of generality, assume the convergence time of the MDB
algorithms in Section~\ref{sec:Algorithms} is the length of a time
slot $T$,\footnote{In practice, the gradient projection algorithms
can only find an approximate optimal solution within a finite
period of time. In this work, we make the idealization that the
exact optimum can be achieved after the convergence period $T$.
Such an assumption simplifies the following analysis while its
loss of precision is small when we take $T$ sufficiently large.}
i.e., at time $\tau = (t+1)T$, the optimal service rate vector for
$\bs U[t]$ is achieved.  For ease of analysis, we further scale
time so that $T = 1$.

We assume a general feasible service rate region. Instead of studying the service rates $(R_{ij}^k(\tau))$, in this
section we focus on the \emph{virtual service rates}. First define the instantaneous virtual service rate of queue
$i^k$ by\footnote{Virtual service rates can be negative, as when a queue's endogenous incoming rate is higher than its
outgoing rate.}
\[ \tilde R_i^k(\tau) = \sum_{j \in {\cal O}(i)} R_{ij}^k(\tau) - \sum_{m \in {\cal I}(i)} R_{mi}^k(\tau). \] Such a
transformation considerably simplifies our subsequent analysis. The vector of virtual service rates over the $t$th slot
is
\[
\tilde{\bs R}[t] = \int_{t}^{t+1} \tilde{\bs R}(\tau) d \tau,
\]
where the integration is taken component-wise. By definition, we have $\tilde R_i^k[t] = \sum_{j \in {\cal O}(i)}
R_{ij}^k[t] - \sum_{m \in {\cal I}(i)} R_{mi}^k[t]$. Therefore, we consider $\tilde{\bs R}[t] = (\tilde R_i^k[t])_{i
\in {\cal N}, k \in {\cal K}}$ induced by $\bs R[t]$.  A virtual service rate vector over a slot $\tilde{\bs R}[t]$ is
feasible if it is induced by a feasible $\bs R[t] \in {\cal C}(\Pi)$. Denote the set of all feasible $ \tilde{\bs
R}[t]$ by ${\cal C}_{\tilde{\bs R}}(\Pi)$. It is straightforward to verify that ${\cal C}_{\tilde{\bs R}}(\Pi)$ is
compact and convex. By Theorem 1 of \cite{paper:NMR05}, the subset of ${\cal C}_{\tilde{\bs R}}(\Pi)$ in the positive
orthant is the stability region of the wireless multi-hop networks with power constraints $\Pi$. For brevity, we denote
${\cal C}_{\tilde{\bs R}}(\Pi)$ by ${\cal C}$ in this section. Finally, the queueing dynamics in
\eqref{eq:QueueDynamic} can be written in vector form as \be\label{eq:GeneralQueueDynamic} \bs U[t+1] \le \left( \bs
U[t] - \tilde{\bs R}[t]  + \bs B[t]\right)^+. \ee

Note that maximizing the MDB objective function \eqref{eq:OptPowerVector} in $\bs R$ over the \emph{feasible
service rate region} ${\cal C}(\Pi)$ is equivalent to maximizing $\bs U[t]' \cdot \tilde{\bs R}$ in
$\tilde{\bs R}$ over the \emph{virtual service rate region} ${\cal C}$. We denote the maximizing $\tilde{\bs
R}$ by $\tilde{\bs R}^*(\bs U[t])$. From now on, we simply call $\tilde{\bs R}$ the service rate vector and
refer to $\tilde{\bs R}^*(\bs U[t])$ as the {\em optimal rate allocation} for queue state $\bs U[t]$.

Recall our discussion of the distributed MDB control algorithms in
the last section. Due to the iterative nature of the algorithms,
the optimal power vector and the optimal rate allocation for a
given queue state can be found only when the algorithms converge.
Therefore in practice, the rate vector solving
\eqref{eq:OptPowerVector} for $(b_{ij}^*[t])$ cannot be applied
instantly at the beginning of the $t$th slot. The actual service
rates $\tilde{\bs R}(\tau), \tau \in \mathbb{R}_+$, are always in
transience, shifting from the previous optimum to the next
optimum.  Thus, the instantaneous rate vector at time $\tau = t$
is $\tilde{\bs R}(t) = \tilde{\bs R}^*(\bs U[t-1])$, and at time
$\tau = t+1$, $\tilde{\bs R}(t+1) = \tilde{\bs R}^*(\bs U[t])$.

\subsection{Lyapunov Drift Criterion}

Following the previous model, the process $\left\{ \left( \bs U[t], \bs U[t-1]\right)\right\}_{t=1}^\infty$
forms a Markov chain. The state $\left( \bs U[t], \bs U[t-1]\right) \triangleq \bs W[t]$ lies in the state
space ${\cal W} =  \mathbb{R}_+^M \times \mathbb{R}_+^M$ where $M$ is the total number of queues. As an
extension of Foster's criterion for a recurrent Markov chain \cite{book:Asm87}, the following condition is
used in studying the stability of stochastic queueing systems \cite{paper:TE92,paper:NMR02}.

\vspace{0.1in}\begin{lemma}\cite{paper:TE92}\label{lma:StabilityCriterion} If there exist a (Lyapunov)
function $V: {\cal W} \mapsto \mathbb{R}_+$, a compact subset ${\cal W}_0 \subset {\cal W}$, and a positive
constant $\varepsilon_0$ such that for all $\bs w \in {\cal W}_0$ \be \mathbb{E}\left[ V(\bs W[t+1]) - V(\bs
W[t]) | \bs W[t] = \bs w \right] < \infty, \label{eq:LyaDrift1} \ee and for all $\bs w \notin {\cal W}_0$ \be
\mathbb{E}\left[ V(\bs W[t+1]) - V(\bs W[t]) | \bs W[t] = \bs w \right] \le -\varepsilon_0,
\label{eq:LyaDrift2} \ee then the Markov chain $\{\bs W[t]\}$ is recurrent.\footnote{For a Markov chain with
continuous state space to be recurrent, the following condition usually is required in addition to those in
Lemma~\ref{lma:StabilityCriterion}: there exists a subset of states which can be visited from any other state
(in a finite number of steps) with positive probability. For $\{\bs W[t]\}$ studied here, the zero state
constitutes such a subset because by assumption $\mathbb{P}(B_i^k[t] = 0) > 0$ for all queues $i^k$.} Hence,
the queueing system is stable in the sense of Definition \ref{def:QueueStable}.
\end{lemma}\vspace{0.1in}


We use the Lyapunov function from \cite{paper:TE96}: \beas V(\bs W[t]) &=& \sum_{k \in {\cal K}}\sum_{i\in
{\cal N}} U_i^k[t]^2 +  (U_i^k[t] - U_i^k[t-1])^2 \\
&=& \| \bs U[t] \|^2 + \| \bs U[t] - \bs U[t-1] \|^2, \eeas where
$\| \cdot \|$ denotes the $L^2$ norm.  Using relation
\eqref{eq:GeneralQueueDynamic}, we derive the following upper
bound on the expected one-step Lyapunov drift conditioned on $\bs
W[t] = \left( \bs u_t, \bs u_{t-1}\right)$: \beas &&
\mathbb{E}\left[ V(\bs W[t+1]) - V(\bs W[t]) | \bs W[t] = \left(
\bs u_t, \bs u_{t-1}\right)\right] \\
&\le& 2 \bs u_t'  \left( \bs a - \tilde{\bs R}[t] \right) + 2 \left( |\bs b| +
\| \tilde{\bs R}[t] \|^2 \right) \\
&&- \| \bs u_t - \bs u_{t-1} \|^2, \eeas where $\bs b$ is the
vector of second moments of the random arrival rates and $|\cdot|$
denotes the $L^1$ norm. The detailed derivation of the above
inequality is left to Appendix~\ref{sec:app}.

Because the distributed power adjustment algorithms in
Section~\ref{sec:Algorithms} increase the objective value $\bs
u_t' \cdot \tilde{\bs R}$ with every iteration from time $t$ to
$t+1$, $\bs u_t' \cdot \tilde{\bs R}(\tau)$ is increasing in $\tau
\in [t, t+1)$ and given $\bs W[t] = \left( \bs u_t, \bs
u_{t-1}\right)$, \beas && \bs u_t' \cdot \tilde{\bs
R}[t] = \int_{t}^{t+1} \bs u_t' \cdot \tilde{\bs R}(\tau) d \tau \ge \\
&& \int_{t}^{t+1} \bs u_t' \cdot \tilde{\bs R}(t) d \tau = \bs u_t' \cdot \tilde{\bs R}(t) = \bs u_t' \cdot
\tilde{\bs R}^* (\bs u_{t-1}). \eeas Also notice that because the second moment vector $\bs b$ is assumed to
be finite and $\tilde{\bs R}[t]$ lies in the bounded region ${\cal C}$, we can find a finite constant
$\lambda$ such that $2 \left( |\bs b| + \| \tilde{\bs R}[t] \|^2 \right) \le \lambda$. Thus, the conditional
expected Lyapunov drift is upper bounded by
\[
2 \bs u_t' \cdot \left( \bs a - \tilde{\bs R}^*(\bs u_{t-1}) \right) - \| \bs u_t - \bs u_{t-1} \|^2 +
\lambda.
\]

Using the above Lyapunov function and the upper bound for the
expected Lyapunov drift, we show the following main result.

\vspace{0.1in}\begin{theorem}\label{thm:StabilityOfAMDB} The iterative MDB policy with convergence time is throughput
optimal, i.e. it stabilizes all arrival processes whose average rate vector $\bs a \in \mathbf{int}~{\cal C}$.
\end{theorem}\vspace{0.1in}

Guided by the Lyapunov drift criterion, the proof aims to find an
$\varepsilon_0 > 0$ and a compact set ${\cal W}_0$ (which may
depend on $\varepsilon_0$) which satisfy the conditions
\eqref{eq:LyaDrift1}-\eqref{eq:LyaDrift2} for any average arrival
rates $\bs a \in \mathbf{int}~{\cal C}$. Note that condition
\eqref{eq:LyaDrift1} is always satisfied since the first and
second moments of arrival rates as well as the service rate vector
are bounded. Now consider the compact region characterized by
\be\label{eq:CompactRegion} {\cal W}_0 = \{ \bs w \in
\mathbb{R}_+^M \times \mathbb{R}_+^M: V(\bs w) \le \Omega \}. \ee
Given $\varepsilon_0
> 0$, we need to specify a finite $\Omega$ and show that when $\bs w[t] = (\bs u_t, \bs
u_{t-1}) \notin {\cal W}_0$, \be\label{eq:DominantInnerProduct} 2 \bs u_t' \cdot \left( \bs a - \tilde{\bs
R}^*(\bs u_{t-1}) \right) - \| \bs u_t - \bs u_{t-1} \|^2 + \lambda \le -\varepsilon_0. \ee

Towards this objective, we devise a geometric method to relate the position of $\bs u_t$ and $\bs u_{t-1}$ in the state
space to the value of the inner product $\bs u_t' \cdot [\bs a - \tilde{\bs R}^*(\bs u_{t-1})]$. In order to reveal the
insight underlying this approach, we first develop the methodology in $\mathbb{R}^2$. The generalization to higher
dimensions as well as the proof for Theorem~\ref{thm:StabilityOfAMDB} can be found in Appendix \ref{app:RM} and
\ref{app:ItMDBOpt}.

\subsection{Geometric Analysis}\label{subsec:Geometric}

In this section, we analyze vectors of arrival rates, service rates, and queue states geometrically. In view of
condition \eqref{eq:DominantInnerProduct}, we characterize a neighborhood around $\bs u_t$ which has the following
properties: if $\bs u_{t-1}$ lies in the neighborhood, then the first term $2 \bs u_t' \cdot ( \bs a - \tilde{\bs
R}^*(\bs u_{t-1}))$ is substantially negative ($\le -\lambda - \varepsilon_0$); if $\bs u_{t-1}$ lies outside the
neighborhood (meaning that $\| \bs u_t - \bs u_{t-1} \|^2$ is relatively large), then the second term $- \| \bs u_t -
\bs u_{t-1} \|^2$ is sufficiently negative for \eqref{eq:DominantInnerProduct} to hold.

We assume an average arrival rate vector $\bs a \in \mathbf{int}~{\cal C}$. There must exist a
point $\bar{\bs a} \in \mathbf{bd}~{\cal C}$, and a positive constant $\varepsilon$ such that $\bs
a + \varepsilon \cdot \bs 1 \le \bar{\bs a}$. Therefore the point $\bs e = \bs a +
\frac{\varepsilon}{2}\cdot \bf 1$ is also in the interior of ${\cal C}$.

Given the current queue state vector $\bs u_t \ge \bs 0$, the hyperplane ${\cal B}_{\bs e}(\bs u_t)
\triangleq \{\bs x: \bs u_t' \cdot \bs x = \bs u_t' \cdot \bs e\}$ is perpendicular to $\bs u_t$ and crosses
the point $\bs e$. The intersection of halfspace ${\cal H}_{\bs e}^+(\bs u_t) \triangleq \{\bs x: \bs u_t'
\cdot \bs x \ge \bs u_t' \cdot \bs e\}$ with ${\cal C}$, denoted by ${\cal C}_{\bs e}^+(\bs u_t)$, is closed
and convex with non-empty interior \cite{book:Egg77}.

\vspace{0.1in}\begin{lemma}\label{lma:Ce+u} For $\bs y \in {\cal
C}_{\bs e}^+(\bs u_t)$, $\bs u_t' \cdot [\bs a - \bs y] \le
-\frac{\varepsilon}{2} \|\bs u_t\|$.
\end{lemma}\vspace{0.1in}

\textit{Proof:} Since $\bs y \in {\cal H}_{\bs e}^+(\bs u_t)$, by definition $\bs u_t' \cdot \bs y \ge \bs
u_t' \cdot \bs e$. Thus,
\beas \bs u_t' \cdot [\bs a - \bs y] &\le& \bs u_t' \cdot [\bs a - \bs e] \\
&=& -\frac{\varepsilon}{2} |\bs u_t| \le -\frac{\varepsilon}{2}
\|\bs u_t\|. \eeas The last inequality follows from $|\bs u_t| \ge
\|\bs u_t\|$ since $\bs u_t \ge \bs 0$. \qed

\subsubsection*{Two-Dimensional Heuristic}

\begin{figure}[h]
\begin{center}
\includegraphics[height = 5cm]{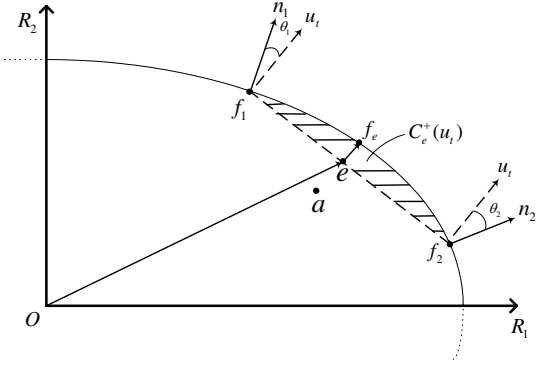}
\caption{The geometry when ${\cal B}_{\bs e}(\bs u_t)$ intersects $\mathbf{bd}~{\cal C}$ at two
different points in $\mathbb{R}_2^+$} \label{fig:TwoDim}
\end{center}
\end{figure}

\begin{figure}[h]
\begin{center}
\includegraphics[height = 5cm]{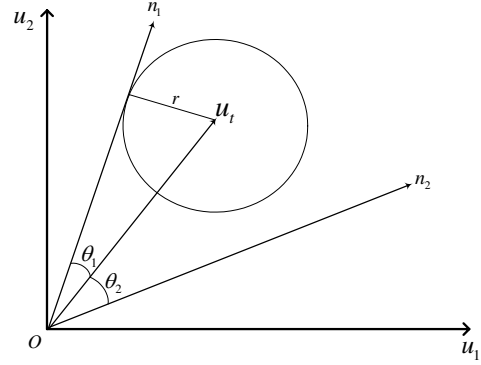}
\caption{The geometry of $\bs u_{t-1}$ lying in the neighborhood
of $\bs u_t$, where $r = \| \bs u_t \| \cdot \alpha(\vec{\bs
u}_t)$. } \label{fig:TwoDimCone}
\end{center}
\end{figure}

Assume there are two queues in the network and index them by $1$ and $2$. In this subsection, all
vectors, hyperplanes, surfaces, etc. are in $\mathbb{R}^2$. The hyperplane ${\cal B}_{\bs e}(\bs
u_t)$ must intersect $\mathbf{bd}~{\cal C}$ at two different points, as illustrated in
Figure~\ref{fig:TwoDim}.
Let the two points be $\bs f_1$ and $\bs f_2$, where $\bs f_1$ is the upper-left one. Denote the hyperplane
(which is a line in $\mathbb{R}^2$) tangent\footnote{The tangent hyperplane contains $\bs f_1$ and defines a
halfspace containing ${\cal C}$.} to ${\cal C}$ at $\bs f_1$ by ${\cal B}_{\bs f_1}(\bs n_1)$, where $\bs
n_1$ is the unit normal vector of the tangent line. Specifically, we require $\bs n_1$ to be pointing outward
from ${\cal C}$. Since ${\cal C}$ is not confined in $\mathbb{R}_+^2$, $\bs f_1$ is not necessarily
nonnegative, and neither is $\bs n_1$. If there exist multiple tangent lines at $\bs f_1$, take $\bs n_1$ to
be any one of them. Let the unit normal vector at $\bs f_2$ be $\bs n_2$, defined in the same manner. Let
\[
\theta_1(\overrightarrow{\bs u_t}) = \arccos (\bs n_1' \cdot
\overrightarrow{\bs u_t}), ~ ~\theta_2(\overrightarrow{\bs u_t}) =
\arccos (\bs n_2' \cdot \overrightarrow{\bs u_t}),
\]
where $\overrightarrow{\bs u_t}$ stands for the normalized vector
of $\bs u_t$. Since $\bs e \in \mathbf{int}~{\cal C}$, $\bs n_1$
and $\bs n_2$ can never be parallel to $\overrightarrow{\bs u_t}$.
Thus,
\[
\bs n_1' \cdot \overrightarrow{\bs u_t} < 1,\quad \quad \bs n_2'
\cdot \overrightarrow{\bs u_t} < 1,
\]
and $\theta_1(\overrightarrow{\bs u_t}) > 0$,
$\theta_2(\overrightarrow{\bs u_t}) > 0$. Moreover,
$\theta_1(\overrightarrow{\bs u_t})$ and
$\theta_2(\overrightarrow{\bs u_t})$ are bounded away from zero
for all $\bs u_t$. To see this, we make use of Figure
\ref{fig:TwoDim} again. The point $\bs f_e$ is on the boundary and
the vector $\bs f_e - \bs e$ is parallel to $\bs u_t$. By simple
geometry, the convexity of the rate region implies $\theta_1(\bs
u_t) \ge \arctan (\|\bs f_e - \bs e\| / \|\bs f_1 - \bs e\|)$.
Because $\bs e$ is an interior point, $\|\bs f_e - \bs e\| \ge \xi
> 0$. Moreover, $\|\bs f_1 - \bs e\| \le D < \infty$ since ${\cal
C}$ is a bounded region. Therefore, $\theta_1(\overrightarrow{\bs
u_t}) \ge \arctan(\xi/D) > 0$. The same is true for
$\theta_2(\overrightarrow{\bs u_t})$. Thus, we can construct a
non-empty cone emanating from the origin sweeping from the
direction of vector $\bs u_t$ clockwise by
$\theta_2(\overrightarrow{\bs u_t})$ and counterclockwise by
$\theta_1(\overrightarrow{\bs u_t})$. Such a cone always contains
$\bs u_t$ in its strict interior.  This is illustrated in
Figure~\ref{fig:TwoDimCone}.

We consider the following two cases. First, if $\|\bs u_t - \bs
u_{t-1}\| / \|\bs u_t\| \le
\sin\left[\min\{\theta_1(\overrightarrow{\bs u_t}),
\theta_2(\overrightarrow{\bs u_t}), \pi/2\}\right] \equiv
\alpha(\overrightarrow{\bs u_t})$, then the pair of points
$\left(\bs u_t, \bs u_{t-1}\right)$ both lie in the cone described
above.
In this case, $\bs u_{t-1}$ is said to be in the {\em neighborhood} of $\bs u_t$. 
See Figure~\ref{fig:TwoDimCone}.

Let $\alpha$ be the infimum of $\alpha(\overrightarrow{\bs u_t})$
over all nonnegative unit vector $\overrightarrow{\bs u_t}$.
Because all $\theta_1(\overrightarrow{\bs u_t})$ and
$\theta_2(\overrightarrow{\bs u_t})$ are strictly positive,
$\alpha$ must be strictly positive. If $\|\bs u_t - \bs u_{t-1}\|
/ \|\bs u_t\| \le \alpha$, $\bs u_{t-1}$ is also in the cone with
$\bs u_t$. In this case, the hyperplane of normal vector $\bs
u_{t-1}$ tangent to the rate region ${\cal C}$ touches
$\mathbf{bd}~{\cal C}$ at $\tilde{\bs R}^*(\bs u_{t-1})$ somewhere
between $\bs f_1$ and $\bs f_2$, i.e., $\tilde{\bs R}^*(\bs
u_{t-1}) \in {\cal C}_{\bs e}^+(\bs u_t)$. By Lemma
\ref{lma:Ce+u}, the inner product $\bs u_t' \cdot [\bs a -
\tilde{\bs R}^*(\bs u_{t-1})] \le -\frac{\varepsilon}{2} \|\bs
u_t\|$. Then for all $\bs w[t]$ such that $V(\bs w[t]) > (1 +
\alpha^2) \cdot (\varepsilon_0 + \lambda)^2 / \varepsilon^2 \equiv
\Omega_1$, $\|\bs u_t\|
> (\varepsilon_0 + \lambda) / \varepsilon $, and therefore \beas && 2 \bs u_t' \cdot \left( \bs a -
\tilde{\bs R}^*(\bs u_{t-1}) \right) - \| \bs u_t - \bs u_{t-1} \|^2
+ \lambda \\
&\le& 2 \bs u_t' \cdot \left( \bs a - \tilde{\bs R}^*(\bs u_{t-1}) \right) + \lambda < - \varepsilon_0, \eeas
which is the desired condition \eqref{eq:DominantInnerProduct}.

If $\|\bs u_t - \bs u_{t-1}\| / \|\bs u_t\| > \alpha$ and assume $\|\bs u_t - \bs u_{t-1}\|^2 =
\omega$, then \beas && 2 \bs u_t' \cdot \left( \bs a - \tilde{\bs R}^*(\bs u_{t-1}) \right) - \|
\bs u_t - \bs u_{t-1} \|^2
+ \lambda \\
&\le& 2 \| \bs u_t \| \| \bs a - \tilde{\bs R}^*(\bs u_{t-1}) \| - \omega + \lambda \\
&<& 2 \sqrt{\omega/ \alpha^2} \sqrt{\lambda / 2} - \omega + \lambda \\
&=& \sqrt{2 \omega \lambda}/ \alpha - \omega + \lambda. \eeas Define \be\label{eq:omega2}\omega_2 =
\inf \{ \omega
> 0: \sqrt{2 \omega \lambda}/ \alpha - \omega + \lambda \le -\varepsilon_0 \}.\ee Then for all $\bs
w[t]$ such that $V(\bs w[t]) > (1 + 1 / \alpha^2) \omega_2 \equiv \Omega_2$, $\|\bs u_t - \bs u_{t-1}\|^2
> \omega_2$ and \eqref{eq:DominantInnerProduct} holds.

Combining the above two cases and letting $\Omega = \max \{
\Omega_1, \Omega_2 \}$, we see that the region specified in
\eqref{eq:CompactRegion} satisfies
Lemma~\ref{lma:StabilityCriterion} and
Theorem~\ref{thm:StabilityOfAMDB} follows.


\section{Numerical Experiments}\label{sec:Simulation}

To assess the practical performance of the node-based distributed
MDB policy in stochastic wireless networks, we conduct the
following simulation to compare the total backlogs resulting from
the same arrival processes under different MDB schemes.

Our scheme iteratively adjusts the transmission powers during a slot to find the optimal rates for the queue state at
the beginning of a slot. As a consequence, the MDB optimization is done with delayed queue state information, the
transmission rates keep changing with time, and the optimal rates are achieved only at the end (beginning) of the
current (next) slot. Recently, Giannoulis et al.~\cite{paper:GTT06} proposed another distributed power control
algorithm to implement the MDB policy in CDMA networks. Instead of converging to the optimal solution for the current
MDB problem, their scheme updates the link powers based on the present queue state only once in a slot. The new queue
state at the beginning of the next slot is used for the subsequent iteration. To highlight the above difference, we
refer to our method as ``iterative MDB with convergence'', and the method studied in~\cite{paper:GTT06} as ``iterative
MDB without convergence''. Both schemes are shown to preserve the throughput optimality of the original MDB policy,
which ideally (instantaneously) finds the optimal transmission rates for the queue state at the beginning of a slot,
and applies them for the whole slot.

For a single run of the experiment, we use a network with $N$ nodes uniformly distributed in a disc of unit radius.
Nodes $i$ and $j$ share a link if their distance $d(i,j)$ is less than $2.5 / \sqrt{N}$, so that the average number of
a node's neighbors remains constant with $N$. The path gain is modeled as $h_{ij} = d(i,j)^{-4}$. The processing gain
of the CDMA system is $K = 10^5$, and the self-interference parameter is $\theta_i = 0.25$. All nodes are subject to
the common total power constraint $\hat P_i = 100$ and AWGN of power $N_i = 0.1$.

\begin{figure}[h]
\begin{center}
\includegraphics[height = 6cm]{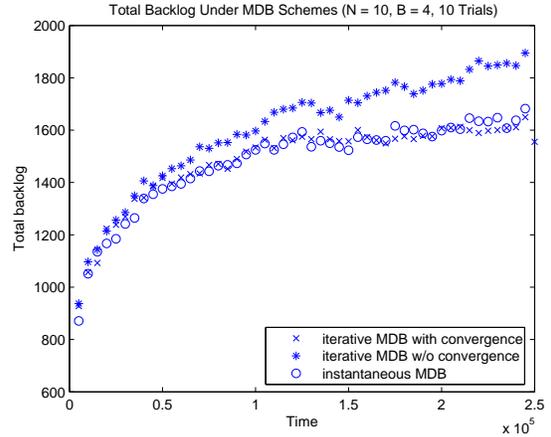}
\caption{Total backlogs under three MDB schemes ($N=10, B=4$).}\label{fig:Sim1}
\end{center}
\end{figure}

\begin{figure}[h]
\begin{center}
\includegraphics[height = 6cm]{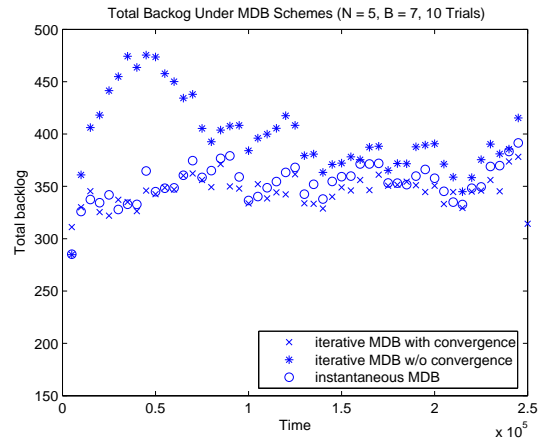}
\caption{Total backlogs under three MDB schemes ($N=5, B=7$).}\label{fig:Sim2}
\end{center}
\end{figure}

Each node is the source node of one session with the destination chosen from the other $N-1$ nodes at random. At the
beginning of every slot, the new arrivals of all $N$ sessions are independent Poisson random variables with the same
parameter $B$. As an approximation, we assume the iterative MDB scheme converges after 50 iterations of the $PA$ and
$PC$ algorithms. The convergence time is taken to be the length of a slot, as in Section~\ref{sec:StabilityOfAsyn}. The
network performance is investigated under each of the MDB schemes with the same set of arrival processes. The total
backlog in the network is recorded after every slot. Figure~\ref{fig:Sim1} shows the backlog curves generated by the
three schemes after averaging $10$ independent runs with the parameters $N = 10$ and $B = 4$.
Figure~\ref{fig:Sim2} reports the result from the experiment with the parameters $N=5$ and $B=7$.
The three methods all manage to stabilize the network queues in the long run. However, the iterative MDB scheme with
convergence and the instantaneous MDB scheme result in lower queue occupancy, hence lower delay, than the iterative MDB
scheme without convergence.

\section{Conclusion}

In this work, we study the distributed implementation of the Maximum Differential Backlog algorithm within
interference-limited CDMA wireless networks with random traffic arrivals. In the first half of the paper, we develop a
set of node-based iterative power allocation and power control algorithms for solving the MDB optimization problem. Our
algorithms are based on the scaled gradient projection method.  We show that the algorithms can solve the MDB
optimization in a distributed manner using low communication overhead. Because these iterative algorithms typically
require non-negligible time to converge, the optimal rate allocation can only be found iteratively over time.   In the
second half of the paper, we analyze the iterative MDB policy with convergence time.  Using a new geometric approach
for analysis of the expected Lyapunov drift, we prove that throughput optimality of the MDB algorithm still holds as
long as the second moments of traffic arrival rates are bounded. The two parts of the paper in conjunction yield a
distributed solution to throughput optimal control of CDMA wireless networks with random traffic arrivals.

\appendix

\subsection{Derivation of Lyapunov Drift}

\label{sec:app}

By definition, the difference of Lyapunov values $V(\bs W[t+1])$ and $V(\bs W[t])$ can be written
as
\beas && V(\bs W[t+1]) - V(\bs W[t]) \\
&=& \| \bs U[t+1] \|^2 - \| \bs U[t] \|^2 + \| \bs U[t+1] - \bs U[t] \|^2 \\
&& - \| \bs U[t] - \bs U[t-1] \|^2 \\
&=& 2 \bs U[t+1]' \cdot \left( \bs U[t+1] - \bs U[t] \right) \\
&& - \| \bs U[t] - \bs U[t-1] \|^2. \eeas

Using relation~\eqref{eq:GeneralQueueDynamic}, we have \beas && \bs U[t+1]' \cdot \left( \bs U[t+1] - \bs U[t] \right) \\
&\le& \left( \left( \bs U[t] - \tilde{\bs R}[t]  + \bs B[t]\right)^+ \right)' \cdot \\
&& \left( \left( \bs U[t] - \tilde{\bs R}[t]
+ \bs B[t]\right)^+ - \bs U[t] \right) \\
&\le& \left(  \bs U[t] - \tilde{\bs R}[t]  + \bs B[t]  \right)' \cdot  \left( \bs B[t] - \tilde{\bs R}[t] \right) \\
&\le& \bs U[t]' \cdot (\bs B[t] - \tilde{\bs R}[t]) + \| \bs B[t] \|^2 + \| \tilde{\bs R}[t] \|^2. \eeas

Therefore, we finally obtain  \beas && V(\bs W[t+1]) - V(\bs W[t]) \\
&\le& 2 \bs U[t]' \cdot (\bs B[t] - \tilde{\bs R}[t]) + 2 \left( \| \bs B[t] \|^2 + \| \tilde{\bs
R}[t] \|^2 \right)\\
&& - \| \bs U[t] - \bs U[t-1] \|^2. \eeas

\subsection{Geometric Analysis in $\mathbb{R}^M$}\label{app:RM}

We now generalize our geometric analysis in Section~\ref{subsec:Geometric} to $M$-dimensional space. We
retain the notation from Section~\ref{subsec:Geometric}.

Analogous to the argument used in the two-dimensional case, we focus on characterizing the neighborhood of $\bs u_t$.

\vspace{0.1in}\begin{lemma}\label{lma:ConeExistence} For any $\bs u_t \ge \bs 0$, there exists a region
${\cal K}(\bs u_t) \subset \mathbb{R}_+^M$ such that

1. $\bs u_t \in {\cal K}(\bs u_t)$;

2. ${\cal K}(\bs u_t)$ has non-empty and convex interior relative to any one-dimensional affine space
containing $\bs u_t$;

3. For all $\bs u_{t-1} \in {\cal K}(\bs u_t)$, the optimal rate vector $\tilde{\bs R}^*(\bs u_{t-1})$ with
respect to $\bs u_{t-1}$ is in ${\cal C}_{\bs e}^+(\bs u_t)$.
\end{lemma}\vspace{0.1in}

%
%

Note that ${\cal K}(\bs u_t)$ is the $M$-dimensional analogue of the circle ${\cal S}(\bs u_t, r)$ of radius
$r$ around ${\bs u}_t$ in Figure~\ref{fig:TwoDimCone}. To facilitate the proof, define the set of feasible
unit incremental vectors around a nonnegative unit vector $\overrightarrow{\bs u}$ as \beas
\Delta_{\overrightarrow{\bs u}} &\triangleq& \left\{\bs \Delta = (\Delta_1,\cdots,\Delta_M): \right. \\
&& \left. \|\bs\Delta\| = 1,~\textrm{and}~\Delta_i^k \ge 0~\textrm{if}~\vec{u}_i^k = 0\right\}.
\eeas  
%

\textit{Proof of Lemma \ref{lma:ConeExistence}:}  Each $\bs \Delta \in \Delta_{\overrightarrow{\bs u_t}}$
spans a one-dimensional affine space containing $\bs u_t$.  It is sufficient to show that given any $\bs
\Delta \in \Delta_{\overrightarrow{\bs u_t}}$, there exists $\bar\delta > 0$ such that for all $\delta \in
[0, \bar\delta]$ and $\bs f \in {\cal C}$ satisfying \be\label{eq:AlphaDelta} (\bs u_t + \delta \bs\Delta)'
\cdot \bs f \ge (\bs u_t + \delta \bs\Delta)' \cdot \bs R, \quad \forall \bs R \in {\cal C}, \ee we have $\bs
f \in {\cal C}_{\bs e}^+(\bs u_t)$.

We prove the claim by construction. We make use of the dominant point $\bar{\bs a}$ of $\bs a$ such that $\bs
a + \varepsilon \cdot \bs 1 \le \bar{\bs a}$ (also $\bs e + \varepsilon / 2 \cdot \bs 1 \le \bar{\bs a}$).
Define the parameter  \be\label{eq:dDelta} d(\bs\Delta) \triangleq \max_{\bs R \in {\cal C}} \bs\Delta' \cdot
(\bs R - \bar{\bs a}), \ee which is at least zero (by setting $\bs R = \bar{\bs a}$ in the objective
function). It is possibly equal to zero, and must be bounded from above, because $\bs\Delta$ is a unit vector
and the optimization region ${\cal C}$ is compact.

Now consider \[ \bar\delta = \frac{\varepsilon \|\bs u_t\|}{2 d(\bs\Delta)}, \] which by the above analysis
is positive. Because ${\cal C}$ is convex and compact, for any $\delta \in [0, \bar\delta]$ there exists at
least one $\bs f$ satisfying \eqref{eq:AlphaDelta}. Picking any one such $\bs f$ and specifically letting
$\bs R = \bar{\bs a}$ on the RHS of \eqref{eq:AlphaDelta}, we have \[ (\bs u_t + \delta \bs\Delta)' \cdot \bs
f \ge (\bs u_t + \delta \bs\Delta)' \cdot \bar{\bs a}.\] By using the inequality
\[
\bs u_t' \cdot \bar{\bs a} \ge \bs u_t' \cdot \bs e + \frac{\varepsilon}{2}|\bs u_t| \ge \bs u_t' \cdot \bs e
+ \frac{\varepsilon}{2}\|\bs u_t\|,
\]
we have \beas
\bs u_t' \cdot \bs f &\ge& \bs u_t' \cdot \bar{\bs a} - \delta \bs\Delta' \cdot (\bs f - \bar{\bs a}) \\
&\ge& \bs u_t' \cdot \bs e + \frac{\varepsilon}{2}\|\bs u_t\| - \delta \bs\Delta' \cdot (\bs f - \bar{\bs a})\\
&\ge& \bs u_t' \cdot \bs e + \frac{\varepsilon}{2}\|\bs u_t\| - \bar\delta \max_{\bs R \in {\cal C}}
\bs\Delta' \cdot (\bs
R - \bar{\bs a}) \\
&=& \bs u_t' \cdot \bs e + \frac{\varepsilon}{2}\|\bs u_t\| - \frac{\varepsilon \|\bs u_t\|}{2 d(\bs\Delta)}
\cdot
d(\bs\Delta) \\
&=& \bs u_t' \cdot \bs e. \eeas Thus, we can conclude that $\bs f \in {\cal C}_{\bs e}^+(\bs u_t)$. Since
$\bs f$ is chosen arbitrarily, the claim at the beginning of the proof is proved.

Finally, define ${\cal K}(\bs u_t)$ as \be\label{eq:Cone} \left\{ \bs u_{t-1} \in \mathbb{R}_+^M: \|\bs
u_{t-1} - \bs u_t \| \le \frac{\varepsilon \|\bs u_t\|}{2 d(\overrightarrow{\bs u_{t-1} - \bs u_t})}\right\},
\ee where $d(\cdot)$ is defined as in \eqref{eq:dDelta}. To accommodate the special case of $\bs u_{t-1} =
\bs u_t$, we define $d(\bs 0) = 0$. It is easily verified that the so-constructed ${\cal K}(\bs u_t)$ is a
valid neighborhood of $\bs u_t$, as required by the lemma. \qed

\subsection{Proof for Theorem \ref{thm:StabilityOfAMDB}}\label{app:ItMDBOpt}

If
\[ \frac{\|\bs u_{t-1} - \bs u_t \|} {\|\bs u_t\|} \le
\frac{\varepsilon}{ 2 \sup_{\overrightarrow{\bs u} \ge \bs 0} d(\overrightarrow{\bs u})} \equiv \alpha,\]
then $\bs u_{t-1} \in {\cal K}(\bs u_t)$ where ${\cal K}(\bs u_t)$ is defined in \eqref{eq:Cone}. In this
case, for all $\bs w[t]$ such that $V(\bs w[t]) > (1 + \alpha^2) \cdot (\varepsilon_0 + \lambda)^2 /
\varepsilon^2 \equiv \Omega_1$, $\|\bs u_t\|
> (\varepsilon_0 + \lambda) / \varepsilon $, and therefore \beas && 2 \bs u_t' \cdot \left( \bs a -
\tilde{\bs R}^*(\bs u_{t-1}) \right) - \| \bs u_t - \bs u_{t-1} \|^2
+ \lambda \\
&\le& 2 \bs u_t' \cdot \left( \bs a - \tilde{\bs R}^*(\bs u_{t-1}) \right) + \lambda < - \varepsilon_0, \eeas
which is the desired condition \eqref{eq:DominantInnerProduct}.

If $\|\bs u_t - \bs u_{t-1}\| / \|\bs u_t\| > \alpha$, define $\omega_2$ as in \eqref{eq:omega2}, then for
all $\bs w[t]$ such that $V(\bs w[t]) > (1 + 1 / \alpha^2) \omega_2 \equiv \Omega_2$, $\|\bs u_t - \bs
u_{t-1}\|^2
> \omega_2$ and \eqref{eq:DominantInnerProduct} holds.

Combining the above two cases and letting $\Omega = \max \{ \Omega_1, \Omega_2 \}$, we see that the region
specified in \eqref{eq:CompactRegion} satisfies Lemma~\ref{lma:StabilityCriterion} and therefore the queueing
system is stable under any average arrival rate vector $\bs a \in \mathbf{int}~{\cal C}$. \qed

\bibliography{DAMDB}
\bibliographystyle{ieeetr}

\end{document}